# HEDGING AND PRODUCTION DECISIONS UNDER UNCERTAINTY: A SURVEY

ABSTRACT: This paper synthesizes and analyzes some important current and recent contributions to the theory of the firm under uncertainty. In so doing, it examines the production and hedging decisions of the competitive firm under a single source and multiple sources of uncertainty.

# 1 Introduction

The main goal of this paper is to synthesize and analyze current and recent significant contributions to the theory of the firm under uncertainty. In so doing, it focuses on the production and hedging decisions of the competitive firm under one source and multiple sources of uncertainty.

Section 2 of the paper investigates the production decisions of the firm in the absence of hedging. The findings are important since a large number of the firms do not have the opportunity to hedge in the futures markets (see Dalal and Alghalith (2008)). This section is arranged into two subsections. The first subsection examines decisions regarding the risky input demand, given a two-input technology and a multiple-input technology. It also discusses the case of multiple risky inputs. The other subsection investigates decisions under joint output price and output/cost uncertainty.

Section 3 examines production and hedging decisions when the firm has the ability to hedge in the futures markets. This section is also divided into subsections based on the nature of hedging: hedging with basis risk, hedging with joint price and output uncertainty, hedging with cost uncertainty, and input hedging. Section 4 provides a summary and concluding remarks.

## 2 Production in the absence of hedging

### 2.1 Cost uncertainty

The literature in uncertainty focused on output price uncertainty (for example, Holthausen (1979) and Anderson and Danthine (1983)), however, cost uncertainty received much less attention. Stewart (1978) and White (1986) assumed that the firm's goal is to decide on the optimal plant size. It therefore have a fixed input, which it will purchase at the current price, so that its price is certain. The variable input will be purchased later and its price is random. They also assumed a fixed output. Viaene and Zilcha (1998) and Alghalith (2007a) provided a more complete analysis by including output as a decision variable. The results are important since hedging in the futures markets is not feasible for many inputs (intermediate goods) especially non-agricultural commodities.

**The two-input case.** A risk-averse, competitive firm produces output using two inputs. The output price is known with certainty, but one of the input prices is uncertain (see Alghalith (2007a)). Profit, $\tilde{\Pi}$, is given by

$$\tilde{\Pi} = pf(x_1, x_2) - \tilde{w}_1 x_1 - w_2 x_2,$$

where $p$ is the output price, $f$ is a neoclassical production function, $x_1$ and $x_2$ are the inputs, and $\tilde{w}_1$ and $w_2$ are their respective prices, $\tilde{w}_1$ is random. The firm maximizes the expected utility of the profit

$$\underset{x_1,x_2}{Max} EU\left(\tilde{\Pi}\right),$$

where $U$ is a strictly concave von Neumann-Morgenstern utility function.

**Proposition 1**. *The introduction of cost uncertainty (i) reduces the optimal risky input (ii) causes the non-risky input to be less (equal to) (greater) than its corresponding level under certainty if the cross partial derivative $f_{12} > (=) (<) 0$.*

**Proof**. From the first-order condition of the maximization, $f_1$ increases and $f_2$ remains unchanged in response to the uncertainty,

$$df_1 = f_{11}dx_1 + f_{12}dx_2 > 0, \tag{1}$$

$$df_2 = f_{22}dx_2 + f_{12}dx_1 = 0, \tag{2}$$

hence

$$df_1 = \left(\frac{f_{11}f_{22} - f_{12}^2}{f_{22}}\right)dx_1 > 0, \tag{3}$$

therefore $dx_1 < 0$ .∎

These results are intuitive since the risky input must fall in response to the risk, whereas the non-risky input reacts to this fall according to the technological relationship between the two inputs. That is, the risk has an indirect impact on the non-risky input through its technological relationship to the risky input. Clearly, the role of the technological relationships was irrelevant to the models of Stewart (1978) and Viaene and Zilcha (1998) since Viaene and Zilcha used a single-input-technology, while Stewart assumed a fixed output.

**Proposition 2**. *When risk aversion increases* (*i*) *the optimal risky input falls* (*ii*) *the optimal non-risky input falls (remains the same) (increases) given* $f_{12} > (=) (<) 0$.

**Proof**. Define $\hat{w}_1$ by $pf_1(\mathbf{x}^*) - \hat{w}_1 = 0$ and let $\tilde{\Pi}$ be the profit when $w_1 = \hat{w}_1$. Also let $\mathbf{x}^*$ and $\bar{\mathbf{x}}$ be the optimal input vector for firm1 and firm 2, respectively. Assume that firm 1 is more risk averse than firm 2. Define $a \equiv EU_2'\left(\tilde{\Pi}\right)(pf_1(\mathbf{x}) - \tilde{w}_1)$; it can be shown that $da > 0$ when risk aversion increases (see Alghalith (2007a). Totally differentiating $a$ and $f_2$, we obtain

$$da = H_{11}dx_1 + H_{12}dx_2 > 0, \tag{4}$$

$$df_2 = f_{22}dx_2 + f_{12}dx_1 = 0. \tag{5}$$

Therefore

$$da = \frac{|H|}{pf_{22}EU_2'\left(\tilde{\Pi}\right)}dx_1 > 0, \tag{6}$$

where $H$ is the Hessian; thus $dx_1 < 0$ when risk aversion increases.∎

The result is also intuitive. The increase in the risk aversion reduces the risky input, while the non-risky input responds according its technological relationship to the risky input.

**The multiple-input case.** With a multiple input price technology, profit is given by

$$\tilde{\Pi} = pf(\mathbf{x}) - \tilde{w}_1 x_1 - \sum_i w_i x_i; \ i \neq 1,$$

where $x_1$ is the risky input (see Alghalith (2007a)).

**Proposition 3**. *The introduction of cost uncertainty (i) reduces the optimal level of the risky input (ii) causes the optimal level of each non-risky input to be less (equal to) (greater) than its corresponding level under certainty if all the inputs are supplemental (independent) (competing) in production (that is, each cross partial derivatives of the production function is positive (zero) (negative)) (iii) reduces the optimal output if all the inputs are supplemental in production.*

**Proof.** *i*) In response to cost uncertainty

$$df_1 = f_{11}dx_1 + \sum_i f_{1i}dx_i > 0, \tag{7}$$

$$df_i = f_{1i}dx_1 + \sum_i f_{ii}dx_i = 0. \tag{8}$$

Thus

$$H \begin{bmatrix} dx_1 \\ \vdots \\ dx_n \end{bmatrix} = \begin{bmatrix} df_1 \\ \vdots \\ df_n \end{bmatrix} = \begin{bmatrix} df_1 \\ \vdots \\ 0 \end{bmatrix}, \tag{9}$$

where $n$ is the number of inputs. The solution yields

$$dx_1 = \frac{|D_1| \, df_1}{|H|}, \tag{10}$$

where $D_1$ is a submatrix of $H$. The sign of $|D_1|$ is the opposite of the sign of $|H|$ (by the second-order condition under certainty) and $df_1 > 0$; thus $dx_1 < 0$ in response to the uncertainty. ∎

(*ii*) From (21),

$$dx_i = \frac{|D_i| \, df_1}{|H|}, \tag{11}$$

where $D_i$ is the corresponding submatrix of $H$; by the properties of the Hessian matrix, $|D_i|$ and $|H|$ have the same (opposite) sign if each cross partial derivative of the production function is negative (positive); also $|D_i| = 0$ if each cross partial derivative is equal to zero. Thus, $dx_i \lesseqgtr 0$ given $\forall f_{ij} \gtreqless 0$; $j = 1, ..., n$ and $i \neq j$. ■

($iii$) The change in output is given by

$$df = f_1 dx_1 + \sum_i f_i dx_i. \qquad (12)$$

If $f_{ij} \geq 0$, then $dx_i \leq 0$, so that $df < 0$ and output clearly falls. ■

**Proposition 4**. *For a homothetic production function, the introduction of input price uncertainty reduces the input ratio, $x_1/x_i$.*

**Proof**. For a homothetic production function, $f_i/f_1 = g(x_1/x_i)$ where $g$ is a monotonic function. Thus $d(x_1/x_i) < 0$ in response to the risk, since $d(f_i/f_1) < 0$. ■

**Proposition 5**. *For a homogeneous production function, the average productivity of $x_1$ increases in response to the introduction of cost uncertainty.*

**Proof**. If the production function is homogeneous of degree $r$, then by

Euler's Theorem

$$\frac{f}{x_1} = \frac{1}{r}\left(f_1 + \sum_i f_i \frac{x_i}{x_1}\right). \tag{13}$$

Cost uncertainty increases $f_1$, leaves $f_i$ unchanged and increases $x_i/x_1$; thus $f/x_1$ increases. ∎

**Multiple cost uncertainty**. With multiple cost uncertainty, profit is given by

$$\tilde{\Pi} = pf(\mathbf{x}) - \sum_i \tilde{w}_i x_i - \sum_j w_j x_j,$$

where $x_i$ is the risky input, $x_j$ is the non-risky input, $w_j$ is the non-random input price, $\tilde{w}_i$ is the random input price. The first-order conditions of the maximization are

$$EU'\left(\tilde{\Pi}^*\right)(pf_i(\mathbf{x}^*) - \tilde{w}_i) = 0; i = 1, ..., m \tag{14}$$

$$(pf_j(\mathbf{x}^*) - w_j) EU'\left(\tilde{\Pi}^*\right) = 0; j = m+1, ..., n \tag{15}$$

where $n$ is the number of inputs and $m$ is the number of risky inputs. It can be shown that the introduction of multiple cost reduces the optimal level of each input if all the inputs are non-competing in production and statistically independent.

## 2.2 Joint output price and output/cost uncertainty

Output is random and it is denoted by $\tilde{q}$. There are two common specifications of $\tilde{q}$ (see Dalal and Alghalith (2008)); additive uncertainty: $\tilde{q} = Y + \theta\tilde{\eta}$, where $\tilde{\eta}$ is random with $E\tilde{\eta} = 0$ and $Var(\tilde{\eta}) = 1$. Multiplicative uncertainty: $\tilde{q} = \tilde{v}Y$ where $\tilde{v}$ is random and $E\tilde{v} = 1$. Thus, in both cases $E\tilde{q} = Y$ is the deterministic output. Costs are known with certainty and are given by a cost function, $c(Y, \mathbf{w})$. Thus, the profit function is

$$\tilde{\Pi} = \tilde{p}\tilde{q} - c(Y, \mathbf{w}),$$

where $\tilde{p}$ is the random output price; the firm seeks to solve

$$\underset{Y}{Max} EU\left(\tilde{\Pi}\right).$$

Using these specifications and a multiple non-risky input technology, Dalal and Alghalith (2008) showed the marginal impact of both output price and output uncertainty on the optimal output. Similar results were obtained by Viaene and Zilcha (1998) using a general production function, where $\tilde{q} = f(x, \tilde{\eta})$ and $\tilde{\eta}$ is a shock. However, they employed a single-input production function; hence profit is given by

$$\tilde{\Pi} = \tilde{p}\tilde{q} - wx.$$

Similar results were obtained by Alghalith (2007a, 2003b) and Viaene and Zilcha (1998) under joint output price and cost uncertainty.

## 2.3 Background risk

The additive form of background risk was the focus of the literature (see, for example, Gollier and Pratt (1996), Quiggin (2003), and Machina (1982)). Other forms of background risk were largely neglected. Exceptions include Franke et al (2006) and Pratt (1988), who examined a multiplicative form. Literature dealing with more general forms of background risk is virtually non-existent. Even the sudies that examined the additive or multiplicative form provided restrictive models and results. They placed restrictions on the functional form, probability distributions, and/or the type of the risk such as undesirable risk. For example, Gollier and Pratt (1996)) and Franke et al (2006) adopted undesirable risk and restricted the functional form of utility.

Another restrictive assumption imposed by the previous models is the statistical independence assumption (that is, the background risk is independent of the other risk). In addition, background risk is normally investigated using choice models (see Quiggin (2003)) and hence the impact of background risk on production decisions is not investigated.

Consequently, the future studies need to relax the independence assumption and adopt a general functional form. They also need to introduce a general form of background risk. Moreover, they should incorporate background risk into theory of the firm.

# 3 Hedging

## 3.1 Hedging with basis risk

Paroush and Wolf (1989) presented a model of output hedging with basis risk, where both output and hedging are the decision variables. They showed that the separation property does not hold in the presence of basis risk and that the presence of the basis risk reduces the optimal output and the optimal hedge. They used a second-order Taylor's approximation of the utility function. The limitations of this approach were discussed by Adam-Muller (2003) and Alghalith (2006b).

Their main results can be extended by using a general utility function and general distributions. Below is a description of the model. The risk-averse firm maximizes the expected utility of the profit

$$\underset{Y,h}{Max} EU\left(\tilde{\Pi}\right).$$

The profit function is specified by

$$\tilde{\Pi} = \tilde{p}Y + (b - \tilde{g})h - c(Y),$$

$$\tilde{p} = \bar{p} + \sigma\tilde{\varepsilon}; E\tilde{\varepsilon} = 0; \tilde{g} = \tilde{p} + \delta\tilde{\xi}; E\tilde{\xi} = 0; E\varepsilon\tilde{\xi} = 0,$$

where $Y$ is output, $h$ is the hedge, $\tilde{p}$ is the random spot price with mean $\bar{p}$ and variance $\sigma^2$, $\tilde{\varepsilon}$ is random, $b$ is the current non-random futures price, $\tilde{g}$ is the random future futures price, $\tilde{\xi}$ is a random term representing the basis risk and independent of $\tilde{\varepsilon}$, $\delta$ is a measure of basis risk[1], and $c(Y)$ is the cost function $(c'(Y) > 0, c''(Y) > 0)$.

---

[1] An increase in $\delta$ means an increase in basis risk. In the absence of basis risk, $\delta = 0$.

The first-order conditions are

$$EU'\left[\tilde{p} - c'(Y)\right] = \left[\bar{p} - c'(Y)\right]EU' + Cov\left(U', \tilde{p}\right) = 0, \qquad (16)$$

$$EU'\left[b - \tilde{g}\right] = \left[b - \bar{p}\right]EU' - \sigma EU'\tilde{\varepsilon} - \delta Cov\left(U', \tilde{\xi}\right) = 0. \qquad (17)$$

We use the superscripts $*$ and $\circ$ to denote the optimal values in the absence and the presence of basis risk, respectively.

**Proposition 6**. $Y^* > Y^\circ$

**Proof**. Adding $(20)$ and $(21)$, we obtain $c'(Y^\circ) = b - \delta Cov\left(U', \tilde{\xi}\right)/EU'$. In the absence of basis risk, $c'(Y^*) = b$ and thus $c'(Y^*) > c'(Y^\circ)$ since $Cov\left(U', \tilde{\xi}\right) > 0$.

**Proposition 7**. $h^* > h^\circ$ (the proof is provided by Alghalith (2005)).

## 3.2 Hedging with joint price and output uncertainty

Without relying on any restrictive assumptions, we show the impact of the output risk on the hedge position of the firm (see Alghalith (2006a)). The standard model specifies the profit function as (see Grant (1985) and Lapan and Moschini (1994), among others)

$$\tilde{\Pi} = \tilde{p}\tilde{v}Y + (b - \tilde{p})h - c(Y),$$

where the variables are defined as before.

**Proposition 8.** $Y^* - h^* = Y^\circ - h^\circ \gtreqless 0$ if $\bar{p} - b \gtreqless 0$.

**Proof.** Without the output uncertainty, the profit function is $\tilde{\Pi} = \tilde{p}Y + (b - \tilde{p})h - c(Y)$. Define $\dot{a} \equiv EU'(\tilde{\Pi})(p - c'(Y))$ and $b \equiv EU'(\tilde{\Pi})(b - p)$; thus $d\dot{a} = H_{YY}dY + H_{Yh}dh$ and $db = H_{hh}dh + H_{Yh}dY$ in response to the introduction of the output uncertainty. Thus

$$dY = \frac{H_{hh}(d\dot{a} + db)}{|H|}, \tag{18}$$

$$dh = \frac{H_{hh}(d\dot{a} + db) - c''(Y)EU'db}{|H|} = dY - \frac{c''(Y)EU'db}{|H|}, \tag{19}$$

thus $dh = dY$ in response to the introduction of output uncertainty.∎

## 3.3 Hedging with cost uncertainty

Though the previous studies examined the impact of cost uncertainty on the risky input or output in the absence of hedging, they hardly examined the impact of cost uncertainty in the context of hedging. Alghalith (2006c) expanded the previous literature by using a multiple-input model.

The profit function is specified by

$$\tilde{\Pi} = \tilde{p}\left(f\left(\mathbf{x}\right) - h\right) + bh - \tilde{w}_1 x_1 - \sum_i w_i x_i.$$

Let $^*$ and $^\circ$ denote the optimal solutions in the absence and the presence of cost uncertainty, respectively.

**Proposition 9**. $f\left(\mathbf{x}^*\right) > f\left(\mathbf{x}^\circ\right); h^* > h^\circ; f\left(\mathbf{x}^*\right) - h^* = f\left(\mathbf{x}^\circ\right) - h^\circ$ if $\tilde{p}$ and $\tilde{w}_1$ are independent (the proof is provided by Alghalith (2006c)). Thus cost uncertainty has an adverse impact on the decision variables.

## 3.4 Input hedging

Paroush and Wolf (1992) investigated a firm which faces input price uncertainty in one input of its two-input production function. They found that the partial cross derivatives of the production function and the market structure of the futures price (upward or downward bias) affect the derived input demand.

Alghalith (2008, 2007b) provided two extensions. First, they generalized

Paroush and Wolf's theorem by using general utility function and general distributions. Second, they added a new theorem that shows the impact of adding basis risk on the optimal hedge.

The firm maximizes the expected utility of the profit

$$\underset{x_1,x_2,h}{Max} EU\left(\tilde{\Pi}\right).$$

The profit function is specified by

$$\tilde{\Pi} = pf(x_1, x_2) - p_1 x_1 - \tilde{p}_2^t x_2 - (b - \tilde{g})h,$$

where $x_1$ and $x_2$ are the two inputs, $p_1$ and $\tilde{p}_2^t$ are their respective prices, $p$ is the non-random output price. The price of $x_2$ is random and given by

$$\tilde{p}_2^t = \bar{p}_2^t + \gamma \tilde{\epsilon}; \quad E\tilde{\epsilon} = 0.$$

There exists futures market for $x_2$; the random future futures price is

$$\tilde{g} = \bar{p}_2^t + \delta \tilde{\xi}; \quad E\tilde{\xi} = 0,$$

where $\tilde{\epsilon}$ and $\tilde{\xi}$ are statistically independent. The amount of the hedged input is denoted by $h$ and $b$ denotes the current futures price.

The first-order conditions are

$$(pf_1 - p_1) EU' = 0, \tag{20}$$

$$\left(pf_2 - \bar{p}_2^t\right) EU' - Cov\left(U', \tilde{p}_2^t\right) = 0, \tag{21}$$

$$\left(\bar{p}_2^t - b\right) EU' + Cov\left(U', \tilde{p}_2^t\right) + \delta Cov\left(U', \tilde{\xi}\right) = 0. \tag{22}$$

. Adding (20) and (21) yields $pf_2 = b - \delta Cov\left(U', \tilde{\xi}\right)/EU'$.

We use °, ˆ, and * to denote the optimal level under complete certainty, uncertainty without basis risk and uncertainty with basis risk, respectively; thus $(x_1^\circ, x_2^\circ)$, $(\hat{x}_1, \hat{x}_2)$, $(x_1^*, x_2^*)$ are the optimal solutions under the above cases.

**Theorem 1.** Assume that $f$ is a neoclassical production function[2] then we obtain the following inputs ranking:

---

[2] For a neoclassical function $f_i > 0$, $f_{ii} < 0$, and $f_{11}f_{22} - f_{12}^2 > 0$.

| | $f_{12} > 0$ | | $f_{12} < 0$ | |
|---|---|---|---|---|
| | $b > \bar{p}_2^t$ | $b < \bar{p}_2^t$ | $b > \bar{p}_2^t$ | $b < \bar{p}_2^t$ |
| $x_1$ | $x_1^* < \hat{x_1} < x_1^\circ$ | $x_1^* < \hat{x_1}; x_1^\circ < \hat{x_1}$ | $x_1^* > \hat{x_1} > x_1^\circ$ | $x_1^* > \hat{x_1}; x_1^\circ > \hat{x_1}$ |
| $x_2$ | $x_2^* < \hat{x_2} < x_2^\circ$ | $x_2^* < \hat{x_2}; x_2^\circ < \hat{x_2}$ | $x_2^* < \hat{x_2} < x_2^\circ$ | $x_2^* < \hat{x_2}; x_2^\circ < \hat{x_2}$ |

**Proof.** Totally differentiating $f_1$ and $f_2$, we obtain

$$df_1 = f_{11}dx_1 + df_{12}dx_2 = 0 \implies dx_1 = -\frac{f_{22}dx_2}{f_{11}}, \qquad (23)$$

$$df_2 = df_{22}dx_2 + df_{12}dx_1, \qquad (24)$$

in response to basis and/or cost risk. Substituting (23) into (24) yields

$$df_2 = \left(\frac{f_{11}f_{22} - f_{12}^2}{f_{11}}\right) dx_2, \qquad (25)$$

thus $df_2$ and $dx_2$ have opposite signs. From the first-order conditions with certainty, absence of basis risk, presence of basis risk, respectively

$$f_2^\circ = \frac{\bar{p}_2^t}{p}, \qquad (26)$$

$$\hat{f_2} = \frac{\bar{p}_2^t}{p} + \frac{Cov(U', \tilde{p}_2^t)}{pEU'} = \frac{b}{p}, \qquad (27)$$

$$f_2^* = \frac{b}{p} - \frac{\delta Cov\left(U', \tilde{\xi}\right)}{pEU'}, \quad (28)$$

combining this with (25) produces the results of Theorem 1.■

**Theorem 2.** The optimal hedge falls when basis risk is introduced; that is, $h^* < \hat{h}$.

**Proof.** Define the sets $A$ and $\sim A$ such that

$$A = \{p_2^t | pf_2^* - p_2^t \geq 0\},$$

$$\sim A = \{p_2^t | pf_2^* - p_2^t \leq 0\}.$$

When $h < x_2$, for any $p_2^t \in A$ and $p_2^{t'} \in \sim A$, we must have Therefore,

$$S \equiv \sup_{p_2^t \in A} U'\left(\tilde{\Pi}^*\right) \leq I \equiv \inf_{p_2^{t'} \in \sim A} U'\left(\tilde{\Pi}^*\right). \quad (29)$$

Since $S$ and $I$ are both positive, there must exist a positive constant $t$ such that

$$\frac{E_{\tilde{\xi}} S}{U'\left(E_{\tilde{\xi}} \tilde{\Pi}^*\right)} \leq t \leq \frac{E_{\tilde{\xi}} I}{U'\left(E_{\tilde{\xi}} \tilde{\Pi}^*\right)}, \quad (30)$$

Thus

$$\left(pf_2^* - p_2^t\right) tU'\left(E_{\tilde{\xi}} \tilde{\Pi}^*\right) \geq \left(pf_2^* - p_2^t\right) E_{\tilde{\xi}} U'\left(\tilde{\Pi}^*\right), \forall p_2^t. \quad (31)$$

Taking expectations with respect to $\tilde{\epsilon}$, we obtain

$$E_{\tilde{\epsilon}}\left(pf_2^* - \tilde{p}_2^t\right)U'\left(E_{\tilde{\xi}}\tilde{\Pi}^*\right) = 0 = E_{\tilde{\epsilon}}\left(pf_2^{\hat{}} - \tilde{p}_2^t\right)U'\left(\tilde{\Pi}^{\hat{}}\right). \tag{32}$$

Let $\alpha \equiv E_{\tilde{\epsilon}}\left(pf_2 - \tilde{p}_2^t\right)U'\left(E_{\tilde{\xi}}\tilde{\Pi}\right)$, then (32) implies that $d\alpha = 0$ when basis risk is added. Totally differentiating $\alpha$ (and holding the parameters constant), we obtain

$$d\alpha = (dx_2 - dh)\, EU''(.)\left(pf_2 - \tilde{p}_2^t\right)^2 + pEU'(.)\, df_2 = 0 \tag{33}$$

and thus $dh < 0$ since $dx_2 < 0$ and $df_2 > 0$. This result holds when $h \geq x_2$ (the proof is similar).∎

This result is intuitive, since the basis risk renders hedging less appealing. Hedging becomes less effective since it will not completely offset the adverse impact of the risks on the inputs. That is, the separation property does not hold. Separation holds if the inputs are independent of the probability distributions and attitudes toward the risks (output decisions are separate from the financial decisions). Note that there is separation in the absence of basis risk.

# 4 Conclusion

In the absence of hedging, the optimal level of the risky input always falls in response to the cost uncertainty. The magnitude of change is determined by the probability distribution of its price. But the change in the optimal level of the non-risky input is determined by the technological relationship between the inputs. Output also falls if the inputs are substitutes. The average productivity of the risky input and the inputs ratio increase for a homogeneous production function.

In the presence of hedging, basis risk/cost uncertainty has a negative impact on both the optimal output and the optimal hedge. Moreover, the introduction of output uncertainty/cost uncertainty does not affect the hedge position of the firm.

This survey can serve as basis for numerous empirical and theoretical future studies in the area. Empirical studies can verify/refute the theoretical propositions of the paper. Also, the theoretical models can be adapted by future studies to accommodate other sources of uncertainty.

# References

[1] Adam-Muller, A. (2003). "An alternative view on cross hedging." EFMA Helsinki Meetings.

[2] Alghalith, M. (2008). "Simultaneous output and output hedging: decision analysis," *Journal of Risk Finance*, **9**, pp 200-205.

[3] Alghalith, M. (2007). New economics of risk and uncertainty: theory and applications. Nova Science Publishers Inc., NY.

[4] Alghalith, M. (2007). "Input hedging: generalizations," Journal of Risk Finance, 2007, **8**, pp 309-312.

[5] Alghalith, M. (2006). "Hedging decisions with price and output uncertainty." *Annals of Finance*, **2**, pp 225-227.

[6] Alghalith, M. (2006). "A note on hedging cost and basis risks." *Economic Modelling*, **23**, pp 534-537.

[7] Alghalith, M. (2006). "A note on output hedging with cost uncertainty," *Managerial and Decision Economics*, **27**, pp 387-389.

[8] Alghalith, M. (2005). "A note on hedging with basis risk-an extension of Paroush and Wolf hedging model," *Economics Bulletin*, **4**, pp 1-6.

[9] Alghalith, M. (2003). "Cost uncertainty with multiple variable inputs." *Atlantic Economic Journal*, **31**, p 290.

[10] Alghalith, M. (2003). "Input demand under multiple uncertainty." University of St Andrews Discussion Paper no. 0304.

[11] Anderson, R. and J. Danthine (1983). "Hedger Diversity in Futures Markets," *Economic Journal*, **93**, pp. 370-389.

[12] Alghalith, M. (2002). "The derived demand and hedging cost uncertainty in the futures markets: note and extensions." University of St Andrews Discussion Paper no. 0210.

[13] Dalal, A. and M. Alghalith (2008). "Production decisions under joint price and production uncertainty," *European Journal of Operational Research*, forthcoming.

[14] Franke, G., Schlesinger, H., Stapleton, R.C. (2006). Multiplicative background risk. *Management Science* **52**, 146-153.

[15] Gollier, C., Pratt, J.W. (1996). Risk vulnerability and the tempering effect of background risk. *Econometrica* **64**, 1109-1124.

[16] Grant, D. (1985). "Theory of the firm with joint price and output risk and a forward market." *American Journal of Agricultural Economics,* **67**, pp 630-635.

[17] Holthasen, D. (1979). "Hedging and the competitive firm under price uncertainty," *The American Economic Review,* **69**, 989-995.

[18] Lapan, H. and G. Moschini (1994). "Futures hedging under price, basis, and production risk." *American Journal of Agricultural Economics*, **76**, pp 456-477.

[19] Machina, M. (1982). Expected utility analysis without the independence axiom. *Econometrica* **50**, 277-323.

[20] Paroush, J. and A.Wolf. (1992). "The derived demand with hedging cost uncertainty in the futures markets." *Economic Journal*, **102**, pp 831-844.

[21] Paroush, J. and A. Wolf (1989). "Production and hedging decisions in the presence of basis Risk." *Journal of Futures Markets*, **9**, pp 547-563.

[22] Pratt, J.W. (1988). Aversion to one risk in the presence of others. *Journal of Risk and Uncertainty* **1**, 395-413.

[23] Quiggin, J. (2003). Background risk in generalized expected utility theory. *Economic Theory* **22**, 607-611.

[24] Stewart, M. (1978). "Factor-price uncertainty with variable proportions." *American Economic Review*, **68**, pp 486-73.

[25] Viaene, J. and I. Zilcha (1998). "The behavior of competitive exporting firms under multiple uncertainty." *International Economic Review*, **39**, pp 591-609.

[26] White, H. (1986). "Uncertainty and factor proportions." *Canadian Journal of Economics*, **19**, pp 814-815.